\documentclass[12pt]{article}
\usepackage{amssymb}
\usepackage{latexsym}
\usepackage{amsmath}
\usepackage[normalsize,hang,bf]{caption}
\usepackage{graphicx}
\input epsf
\newcommand{\cg}{{\cal G}{}}
\newcommand{\cj}{{\cal J}{}}
\newcommand{\co}{{\cal O}{}}
\newcommand{\cf}{{\cal F}{}}

\newcommand{\cu}{{\cal U}{}}
\newcommand{\ls}{{\frak{sl}(2,\mathbb{R})}{}}
\newcommand{\SL}{\mbox{SL}(2,\mathbb{R})}
\newcommand{\h}{\,{\bf H}}
\renewcommand{\t}{\,{\bf T}}
\newcommand{\e}{\,{\bf E}}
\newcommand{\f}{\,{\bf F}}
\newcommand{\s}{\,{\bf S}}
\newcommand{\n}{\,{\bf N}}
\newcommand{\K}{\,{\bf K}}
\newcommand{\x}{\,{\bf X}}
\newcommand{\y}{\,{\bf Y}}
\newcommand{\z}{\,{\bf Z}}

\newcommand{\LL}{\,{\bf L}}

\newcommand{\bz}{\,{\bf z}}
\newcommand{\be}{\,{\bf e}}
\newcommand{\ZZ}{\mathbb Z}
\newcommand{\RR}{\mathbb R}
\newcommand{\Id}{\mbox{\rm \bf 1\hspace{-1.2 mm}I}}

\newcommand{\ddto}{{\left.\frac{d}{dt}\right|_0}}
\newcommand{\tr}{\mbox{\rm Tr\,}}

\begin{document}

\begin{center} {\large Regular Poisson structures
on massive non-rotating BTZ black holes}\\

\vspace{.5cm}
P.~Bieliavsky${}^{a,}$\footnote{
E-mail : pbiel@ulb.ac.be}
M.~Rooman${}^{b,}$\footnote{
E-mail : mrooman@ulb.ac.be, FNRS Research Director}
and Ph.~Spindel${}^{c,}$\footnote{ E-mail :
spindel@umh.ac.be}\\

\vspace{.3cm}  {${}^a${\it Service de G\'eom\'etrie diff\'erentielle}}\\
{\it Universit\'e
Libre de Bruxelles, Campus Plaine, C.P. 218}\\ {\it Boulevard du Triomphe,
B-1050 Bruxelles, Belgium}\\
\vspace{.3cm}  {${}^b${\it Service de Physique th\'eorique}}\\
{\it Universit\'e
Libre de Bruxelles, Campus Plaine, C.P.225}\\ {\it Boulevard du Triomphe,
B-1050 Bruxelles, Belgium}\\
\vspace{.2cm} {${}^c${\it
M\'ecanique et Gravitation}}\\ {\it Universit\'e de Mons-Hainaut, 20
Place du Parc}\\
{\it 7000 Mons, Belgium}\\ \end{center}
\vspace{.1cm}

\begin{abstract}
We revisit the non-rotating massive BTZ black hole within a
pseudo-Riemannian symmetric space context.  Using classical symmetric space
techniques we find that every such space intrinsically carries a
regular Poisson structure whose symplectic leaves are para-hermitian
symmetric surfaces.  We also obtain a global expression of the metric
yielding a dynamical description of the black hole from its initial to
its final singularity.
\end{abstract}
\noindent
{\it PACS\/: 97.60.Lf, 02.20.-a, 04.60.Kz}\\
{\it Keywords}: anti de Sitter, BTZ metrics, bi-quotient spaces,
pseudo-Riemannian symmetric spaces.

\section{Introduction}
It is well known that black hole solutions in (2+1) dimensions can be
obtained by performing identifications in $AdS_3$ space by means of
a discrete subgroup of isometries which is isomorphic to $\ZZ$,
as shown in a seminal paper by Ba\~nados, Henneaux, Teitelboim
and Zanelli (BHTZ) \cite{BTZ}.

In the present article, we investigate the BHTZ construction
in the case $J=0,M>0$ within a Lie theoretical and symmetric spaces
framework.  First, we interpret the BHTZ $\ZZ$-action as the
restriction to a one (integral) parameter subgroup of $\SL$ of a
bi-action (denoted hereafter $\Sigma^{\sigma}$) of the whole $\SL$ on
itself associated to an external automorphism $\sigma$ of $\SL$.  The
orbits of the latter are two dimensional and all isomorphic to
one another as homogeneous spaces.  The induced foliation of the group
manifold $G=\SL$ turns out to be a (globally) trivial fibration of
$G=\mbox{AdS}_{3}$ with Lorentzian symmetric fibers.  Obviously, each
fiber is stable by the BHTZ $\ZZ$-action which, moreover, pointwisely
fixes the transverse direction.  

Fibrations of symmetric spaces with
symmetric fibers have been extensively studied for decades, for a
general theory see e.g. \cite{Wolf}. We observe here that the
BHTZ construction (in the massive non-rotating case) appears as a
particular case: a (trivial) fibration of $\mbox{AdS}_{3}$ by
$\mbox{AdS}_{2}$ spaces over $\RR$ with a vertical action of $\ZZ$.
The $\ZZ$-action is not properly discontinuous on the vertical fiber.
In order to guarantee a smooth manifold structure on the (two dimensional)
quotient, one thus needs to restrict the $\ZZ$-action to a (maximal stable)
proper open domain of the fiber, say $\cu^{0}$.  This is what is responsible
for the black hole structure of the three dimensional quotient, which
in this setting is simply globally given by
$(\ZZ\backslash\cu^{0})\times\RR$.  Note that in the BHTZ context,
$\mbox{AdS}_{2}$ spaces as orbits of an external bi-action have
already been considered in \cite{BaPe}.  
Such an $\mbox{AdS}_{2}$
space is the (simplest) two dimensional version of causal symmetric
spaces of the so-called Cayley type defined through the class of
para-hermitian spaces by Kaneyuki in the eighties (see e.g.
\cite{KANEYUKI}).  

This leads to the second aspect of the paper,
namely Poisson structures on BHTZ spaces.  A symmetric space of Cayley
type is not (invariantly) pseudo-K\"{a}hler.  Nevertheless, it always
carries an invariant symplectic structure, defined for instance by
contraction of its para-complex structure (of course, in the two
dimensional case, the resulting symplectic structure is nothing else
than the area element, but the para-complex structure persists).  The
para-complex structure being invariant under the action of the
automorphism group of the Cayley symmetric space, it passes through
the quotient when considering any (even local) pseudo-Riemannian
covering defined on any open domain of the Cayley symmetric space at
hand.  Of course, the same holds for the symplectic structure as well. 
The identification of the BHTZ black hole as a product manifold
$(\ZZ\backslash\cu^{0})\times\RR$ therefore provides a foliated
symplectic structure on the black hole.  In other words, the black
hole carries a canonical Poisson structure whose characteristic
foliation coincides with the above mentioned fiber structure.  Note
that the foliated two-form cannot be seen as the restriction of a
closed two-form on $\mbox{AdS}_{3}$.

The paper is organized as follows.
We first recall, in Section 2, some basic results
about the $AdS_3$ space viewed as the $\SL$ group manifold and fix
conventions and notations.
We also recover the classification of one-parameter groups of isometries of
$AdS_3$, appearing quite simply in this context.  In
Section 3, we show that non rotating BTZ black holes can be viewed as
bi-quotients of $AdS_3$.  Two types of bi-quotients are possible.  The
first, considered in \cite{Brill}, leads to a singular foliation.  The
second, which we shall deal with in Section 4, relies on a twisted
bi-action.  It provides a simple foliation by Lorentzian leaves.  As a
byproduct we obtain a global expression of the metric, providing a
simple dynamical description of the evolution of the black hole.  In
section 5, we define a regular Poisson structure on the BTZ black
hole.  We end the paper, in Section 6, with a short discussion of our
results and their perspectives.

\section{$AdS_3$ space as $\SL$ group manifold}
\subsection*{Metric and conformal structure}
The Lorentzian space $AdS_3$ is defined as the solution of Einstein
equations in $2+1$ dimension with negative cosmological constant. It is
isometric to the Lorentzian group manifold $(G,\beta)$ underlying the
simple real Lie group
\begin{equation}
\label{matbz}
G=\SL=\{\bz =
\left(
\begin{array}{cc}
u+x & y+t\\
y-t & u-x
\end{array}
\right)| x,y,u,t\in\mathbb{R},\ \det \bz =1
\} \qquad ,
\end{equation}
endowed with its Killing metric $\beta$. It can also be viewed as the
quadric $t^2+u^2-x^2-y^2=1$ in the four dimensional flat space $\mathbb{R}^4$
endowed with a ultrahyperbolic metric of signature $(2,2)$:
\begin{equation}
ds^2=-du^2-dt^2+dx^2+dy^2\qquad .
\end{equation}
Particularly useful global coordinates (see Figs 1, 2 and 3) are 
obtained by the
global (up to trivial polar singularities) parametrization:
\begin{eqnarray}
u=\cosh\xi\,\cos\lambda\qquad ,& \hspace{10mm} & t=\cosh\xi\,\sin\lambda
\qquad ,\nonumber\\ x=\sinh\xi\,\cos\varphi \qquad ,& \hspace{10mm} &
y=\sinh\xi\,\sin\varphi
\qquad ,
\end{eqnarray}
which leads to the metric expression
\begin{equation}
\label{metric}
ds^2=\beta=-\cosh^2\xi\, d\lambda^2 + d\xi^2 + \sinh^2\xi\,d\varphi^2\qquad.
\end{equation}
The change of radial coordinate:
\begin{equation}
\tanh\xi=\sin r \hspace{10mm} r \in[0,\pi/2[ \qquad ,
\end{equation}
yields a conformal embedding of $AdS_3$ in the Einstein static space
\cite{HE}:
\begin{equation}\label{he} ds^2=\frac 1{\cos^2 r }\left( -d\lambda^2+d r
^2+\sin^2 r \,d \varphi ^2\right) \qquad .
\end{equation}

BTZ black hole solutions, characterized by their mass $M$ and their
angular momentum $J$, can be obtained by
performing identifications in $AdS_3$ space by means of the action of
  a discrete subgroup of
isometries isomorphic to $\ZZ$ and generated by $\exp{(\frac \pi
2 \Xi)}$, where $\Xi$ is a Killing vector field \cite{BTZ}.
Here, we focus on spinless ($J=0$) BTZ solutions and show, using a group
theoretical approach, that they
admit a canonical global foliation, which allows to construct global
coordinates.

\subsection*{$\SL$: Lie algebra and automorphisms}

The Lie algebra of the group $G$:
\begin{equation}
\cg=\ls=\{\left(\begin{array}{cc} z^H & z^E \\
z^F & -z^H
\end{array}\right)=: z^H\h+z^E\e+z^F\f\} \qquad ,
\end{equation}
is expressed in terms of the generators $\{\h , \e, \f \}$ satisfying
   the commutation relations:
\begin{equation}
\label{comrel}
[\h ,\e]=2\e \quad , \quad [\h,\f]=-2\f \quad ,  \quad [\e,\f]=\h
\quad .
\end{equation}
By identifying $\cal G$ with the tangent space of $G$  at the neutral element
$\be$, the Killing metric $\beta_{\be}$ is defined as:
\begin{equation}
\beta_{\be}(\x,\y):= {\frac 1 8} \tr(ad(\x) ad(\y)) =
{\frac 1 2} \tr({\bf XY}) \qquad ,
\end{equation}
where $ad(\x) \z = [\x,\z]$ and where $\x$, $\y$
and $\z$ are elements of $\cal G$.
There is an isomorphism between $\cal G$ and
Minkowski space in 2+1 dimensions: $\beta_{\be}({\h},{\h})=1$,
$\beta_{\be}({\e},{\f})=1/2$, the other metric coefficients being
zero. Thus {\h} is space-like, whereas {\e} and {\f}
are light-like.

From the Lie algebra commutation relations (\ref {comrel}) we have
that the automorphism group of $\ls$ is isomorphic to the
three dimensional Lorentz group $SO(2,1)\equiv L_+^{\uparrow}(2,1)\cup
L_-^{\downarrow}(2,1)$.  Among these transformations, those belonging
to $L_+^{\uparrow}(2,1)$ correspond to internal automorphisms.  In
contrast, the automorphisms belonging to $L_-^{\downarrow}(2,1)$ need
the introduction of an external automorphism $\sigma$ that, at the
level of the Lie algebra, can be chosen as:
\begin{equation}
\label{defsig}
\sigma(\h)=\h,\ \sigma(\e)= -\e ,\ \sigma(\f) = -\f \qquad .
\end{equation}
Viewing $G$ as a subgroup of ${GL(2,\mathbb{R})}$, one may express
the automorphism $\sigma$ as:
\begin{equation}
\sigma=Ad(\h) \qquad \mbox{where} \qquad \h=\left(
\begin{array}{cc}
1 & 0\\
0 & -1
\end{array}
\right)
  \in GL(2,\mathbb{R}) \qquad .
\end{equation}

\subsection*{Isometry group of $AdS_3$}
The group Iso$(AdS_3)$ is isomorphic to $O(2,2)$ and is thus
constituted by four
connected parts: $O_{++}(2,2)\,\cup\, O_{+-}(2,2)\,\cup\,
O_{-+}(2,2)\,\cup\,
O_{--}(2,2)$. The identity component Iso$_0(AdS_3)$, corresponding to
$O_{++}(2,2)$, is locally isomorphic to $\SL \times \SL$, and we may
therefore represent its action by
using the matrix parametrization (\ref{matbz}) of the points of $AdS_3$, as:
\begin{equation}
  \bz \mapsto g'\, \bz \, g^{-1} \quad,\quad g',g\in \SL\qquad .
\end{equation}
In order to describe the full action of Iso$(AdS_3)$ we have to 
consider in addition the
following two transformations $\cal P$ and $\cal T$ defined by:
\begin{eqnarray}
{\cal P}(\bz)&=&\left
(\begin{array}{cc}u-x&y+t\\y-t&u+x\end{array}\right)\quad,\label{revx}\\
    {\cal T}(\bz)&=&\left
(\begin{array}{cc}u+x&y-t\\y+t&u-x\end{array}\right)\quad,\label{revt}
\end{eqnarray}
which allow to reach the components $O_{+-}(2,2)$,
$O_{-+}(2,2)$ and $O_{--}(2,2)$.

The Lie algebra of Iso$(AdS_3)$
is isomorphic to the
direct product of Lie algebras
$\cg\oplus\cg$. Its action in terms of vector fields is given by:
\begin{equation}
(\x,\y) \mapsto \overline{\x}-\underline{\y}\qquad,
\end{equation}
where $\overline{\x}$ (resp.
$\underline{\y}$) denotes the right-invariant (resp. left-invariant) vector
field on the Lie group $G$ associated to the element $\x$ (resp.
$\y$) of its Lie algebra $\cg$. This means that, at point $\bz$ of $G$,
$\overline{\x}(\bz)=\ddto \exp(t \x) \bz $
and $\underline{\y}(\bz)=\ddto \bz \exp(t \y)$. Note that, as we
deal with a matrix group, this simply reduces to
$\overline{\x}(\bz)=\x \bz$ and
$\underline{\y}(\bz)=\bz \y$.

\subsection*{One-parameter subgroup classification of ${Iso_0(AdS_3)}$}
The classification, up to conjugation, of the one-parameter subgroups of
Iso$_0(AdS_3)$ is straightforward to obtain in this framework.
First recall that two subgroups $H_1$ and $H_2$ of Iso$_0(AdS_3)$ are
conjugated if and only if there exists $ g\in \mbox{Iso}(AdS_3)$
such that $H_1= g\; H_2\; g^{-1}$.
A one-parameter isometry subgroup is specified by two $\SL$ one-parameter
subgroups
$(H_L,H_R)$ defined up to the combination of transformations by $\cal P$
and $\cal T$ and the equivalence relation:
\begin{equation}
\label{equiv}
(H_L,H_R)\equiv (H'_L,H'_R)\ \quad \leftrightarrow\quad g_L,g_R \in
SL(2,\mathbb R) \ {\rm{such\ that}} \ H_i=g_i H'_i g_i^{-1} \quad ,
\end{equation}
where $i=L,R$. The generators
of the $\SL$ one-parameter subgroups are always given, after a suitable
$O_{++}(2,2)$ (Lorentz) transformation, by one of the four generators:
$\h, \e, \f$
and
$\t=\e-\f$. Moreover the $\cal P$ and $\cal T$
transformations interchange
the generators of the left ($L$) and right ($R$) $\SL$ subgroups
according to the rules:
\begin{equation}
\label{classif}
{\cal P}:\left(\begin{array}{c}\h _L\\ \e _L\\ \f
_L \end{array}\right)\leftrightarrow\left(\begin{array}{c}-\h _R\\ \e _R\\ \f
_R \end{array}\right)\qquad{\cal T}:\left(\begin{array}{c}\h _L\\ \e _L\\ \f
_L \end{array}\right)\leftrightarrow\left(\begin{array}{c} \h _R\\ \f _R\\ \e
_R \end{array}\right)\qquad .
\end{equation}
The one-parameter subgroups of Iso$_0(AdS_3)$ are thus all equivalent to
those generated
by the six  different choices of $\ls$ generators (labelled according to the
classification given in \cite{BTZ}):
\begin{eqnarray}
I_a&&(a \t_L\, , \,   b \h_R)\sim (b \h_L\, , \,   a \t_R)\qquad
\; (a > 0,\ b \geq 0)\qquad ,\nonumber\\
I_b&&(a\h_L\,,\, b\h_R)\sim ( b \h_L\, , \,  a \h_R)\qquad
(a > 0,\ b\geq 0) \qquad ,\nonumber\\
I_c&&(a \t_L\, , \,  b \t_R)\sim (b \t_L\, , \, a \t_R)\qquad \
(a>0, b\neq 0)\; \qquad ,\nonumber\\
II_a&&(a \h_L\, , \,\e_R)\sim (\e_L\, , \, a \h_R)\sim (\f_L\, , \,
a \h_R)\sim (a \h_L,\f_R)\  (a\ge 0) \ , \nonumber\\
II_b&&(a \t_L\, , \,\e_R)\sim (\e_L\, , \, a \t_R)\sim (\f_L\, , \,
-a \t_R)\sim
(-a \t_L\, , \,\f_R)\  (a\ge 0) \ , \nonumber\\
III^-&&(\e_L\, , \,\e_R)\sim(\f_L\, , \,\f_R)\qquad,\nonumber\\
III^+&&(\e_L\, , \,-\e_R)\sim(\f_L\, , \,-\f_R)\qquad.
\end{eqnarray}
The restrictions on the sign of coefficients $a$ and $b$ are obtained
as follows. First  their
overall value can always be changed by considering the inverse
subgroup instead
of the subgroup. Furthermore the sign of the coefficient of an
$\h$ generator can be changed as will, using a rotation around the
$\t$ axis in
$\ls$, whereas Lorentz transformations preserve the orientation
of the time-like generator
$\t$ and allows only to scale by a positive factor the light-like
generators $\e$ or $\f$. In the following, we shall extend this
classification to Killing vector fields.

\section{BTZ as bi-quotient space}
Let $G=\SL$ and $H$ be an integral one parameter subgroup of $G$ i.e.
a subgroup of the form $H=\{\exp(n X)\}_{n\in\mathbb{Z}}$ where $X\in\sl$
is some fixed element of the Lie algebra. With this data, one can define
different types of quotient spaces.  First, one can consider the left 
($H\backslash G$)
(resp. right ($G/H$)) coset space, where the subgroup $H$ acts on $G$
by left (resp. right) translations. These give rise to the so called
self- (resp. anti-self-) dual solutions studied in \cite{CH}.
Note that these do not provide black hole solutions.  Indeed, the
element $X$ being conjugated to a multiple of either $\t,\h$ or $\e$
(see preceding section), one may assume the subgroup $H$ to be an
integral subgroup of either $K=SO(2)=\exp(\mathbb{R}\t)$,
$A=SO(1,1)=\exp(\mathbb{R}\h)$ or $N=\exp(\mathbb{R}\e)$.  Therefore,
the Iwasawa decomposition $G=KAN=KNA$ of $G$ \cite{Hel} tells us that
in any case the quotient space has a structure of smooth manifold and
that $G\to H\backslash G$ (resp. $G\to G/H$) is a (pseudo) Riemannian
covering (w.r.t the Killing metric on $G$).  Projecting the geodesics, one
therefore sees that the quotient space is geodesically complete since
$G=AdS_{3}$ is so.  

To obtain black hole solutions, one may consider a
different kind of quotient, defined through a {\em bi-action}.  Let
$G$ be a group and $H_L$ and $H_R$ subgroups of $G$.  By definition, the
bi-action of the subgroups $H_L$ and $H_R$ on the group $G$ is defined
by the mapping:
\begin{eqnarray}
&((H_L,H_R), G)\longrightarrow & G \nonumber\\
& ((k_L,k_R),\bz)\mapsto& k_L \ \bz \ k_R^{-1} \qquad .
\end{eqnarray}
This mapping defines
an equivalence relation on $G$ and one defines the {\em bi-quotient} as
the quotient of $G$ by this relation.  Defining $k_L:=\exp(t \K _L)$
and $k_R:=\exp(t \K _R)$, with $(\K_L,\K_R)\in \cg \oplus \cg$, the
Killing vector field associated to this bi-action is given by:
\begin{equation}
\Xi =\overline{\K_L}-\underline{\K_R} \qquad .
\end{equation}
Its mass $M(\Xi)$ and its angular momentum
$J(\Xi)$ are respectively defined as
\begin{eqnarray}
M(\Xi)&=& \frac 1 2 \left ( \|\K_L\|^2+\|\K_R\|^2 \right )\qquad,\label{mass}\\
J(\Xi)&=&\frac 1 2  \left ( \|\K_L\|^2-\|\K_R\|^2 \right )\qquad,\label{angm}
\end{eqnarray}
where $\|\x\|^2$ is the Killing norm $\beta_e (\x,\x)$.
In the case of a spinless BTZ black hole, $\Xi$ must satisfy the
following conditions:
\begin{enumerate}
\item[(i)] $M(\Xi)\geq 0$,
\item[(ii)] $J(\Xi)=0$,
\item[(iii)] there exists an element $\bz \in AdS_3$ such that
$\beta_{\bz}(\Xi,\Xi)>0$.
\end{enumerate}
The first condition is necessary in order to avoid naked
singularities and the third
to avoid closed time-like curves passing through every point.

For these conditions to be satisfied, the generators $\K_R$
and $\K_L$
must be elements of the same
space-like or light-like sphere in $({\cal G}, \beta_e)$
and there must be a region in $AdS_3$
where $\| \Xi \| ^2 > 0$.
The bi-action is thus defined by a single subgroup $H \subset G$
and an automorphism $\alpha$ of $G$:
\begin{eqnarray}
\label{tau}
&(H, G)\longrightarrow& G \nonumber\\
& (h,g)\mapsto& \Sigma ^\alpha_h (g) :=hg\alpha(h^{-1}) \qquad .
\end{eqnarray}
The automorphism $\alpha$ may be either internal (i.e. conjugated to the
identity) or external (i.e. conjugated to $\sigma$ defined in eq.
(\ref{defsig}))
when $M(\Xi)> 0$, but it is necessarily internal for $M(\Xi)=0$.
Indeed, in the latter case, $\Xi$ must be chosen in class $III^+$
of the Iso$(AdS_3)$ subgroup classification (\ref{classif})
(class $III^-$ generates everywhere closed time-like curves), and the only
automorphisms that keep its generators unchanged are internal.
In contradistinction, when $M(\Xi)>0$, $\Xi$ is
in class $I_b$, which is preserved as well by internal as by external
automorphisms.
In this case,
we may choose without loss of generality:
\begin{equation}
\label{genBHTZ}
\Xi=a\ ( \overline{\h} - \underline{\h}) \qquad {\rm and} \qquad
\alpha=Id \qquad {\rm or} \qquad \alpha=\sigma \qquad ,
\end{equation}
where $\sigma$ is defined in eq. (\ref{defsig}).
The minus sign in front of $\underline{\h}$ is chosen in order to fix the
identity element of $G$.
This choice differs from the one made in \cite{BTZ},
but is of course equivalent.
The parameter $a$ is
related to the mass of the black hole by the
relation (\ref{mass}), i.e. $a=\sqrt{M}$.
Hereafter we shall call the BHTZ subgroup the
one parameter isometry subgroup of $AdS_3$ generated by eq. (\ref{genBHTZ}).

\section{Construction of a global BTZ metric via $AdS_3$ foliation}
Consider an open and connected domain ${\cal U}
\subseteq AdS_3$, where
$ \forall \bz \in {\cal U} : \| \Xi \| ^2 _{\bz} >0$,
and which is invariant under the action of $\mathbb Z$ on
$AdS_3$  given by:
\begin{equation}
\label{zet}
\ZZ \times AdS_3 \rightarrow AdS_3  :
(n,\bz) \rightarrow \psi_n (\bz)  := \exp{_{Iso({\rm AdS}_3)}(n \, \pi \, \Xi)}
\qquad  ,
\end{equation}
which corresponds to the value of the solution for $\rho = n$
of the first-order differential equation
\begin{equation}
\frac {d \psi_{\rho}(\bz)} {d \rho} =  \Xi_{ \psi_{\rho}(\bz)}
\qquad {\rm with} \qquad
\psi_0(\bz)= \bz \qquad .
\end{equation}

Let us moreover assume
that the space of orbits $ {\cal U}/ \ZZ$  admits the structure of
a differential manifold whose underlying topological structure is the quotient
topology, and that $\cal U$ is maximal for these properties.
A $J=0$ BTZ solution can then be viewed as a space of orbits $ {\cal U}/
\ZZ$. To obtain a geometrical description of these spaces, we
construct a foliation that is stable with respect
to the action of $\mathbb Z$. The advantage of such a foliation is that it
provides a family of equivalent (isomorphic) 2-dimensional surfaces,
which are easy to study.

The foliation that we want to construct has to be related to
$\SL$, so that we can perform the quotient by $\ZZ$, or possibly by other
Fuchsian groups, for example to obtain black hole solutions with
more complex topologies.
This procedure has been initiated in \cite{Brill},
whose authors first chose a
space-like surface
${\cal B}_0$ that is
totally geodesic and then propagated the action of group (see Fig. 1).
This procedure is equivalent to choosing
in eq. (\ref{tau}) $\alpha=Id$. It however suffers from several defects.
In $AdS$ spaces the
negative cosmological constant tends to focalize geodesics. This leads to
singularities and does not allow a global description of the
solution, even in the
simplest case. The orbits so obtained change from Riemannian to
Lorentzian beyond the Cauchy horizons of ${\cal B}_0$.

We shall, on the contrary, make use from the beginning of a Lorentzian
surface
instead of a space-like one, choosing in eq. (\ref{tau}) $\alpha=\sigma$.
In this way, we shall construct
time-like surfaces that are
stable with respect to the action of an $\SL$ group, subgroup of
$SO(2,2)$.
This requires
to restrict ourselves to $J=0$ because the one parameter subgroup used
to perform
identifications leading to rotating black holes cannot be embedded in a $\SL$
subgroup of $SO(2,2)$. Indeed, if we start with a generator
$(a \h_L , b \h_R)$, with $a \neq 0 \neq b$, the requirement that
they satisfy the $\ls$ commutation relations
automatically imposes $a=\pm b$ and thus $J=0$.

The foliation we shall obtain will be
everywhere regular and the intersection of a
causally safe region with the orbits are all isomorphic. This will
allow us to obtain global coordinates on
$J=0,\ M> 0$ black-holes that clearly show the dynamical structure of
the solution in terms of a time evolution
of a moduli defining non trivial global space-like sections.

Non-rotating BTZ black holes can be
obtained as the quotient of $\SL$ by the bi-action $\Sigma$ defined in
eq. (\ref{tau})
restricted to $\ZZ$ (see eq. (\ref{zet})).
Indeed, using eq.(\ref{defsig}), we observe that:
\begin{equation}
\label{defpsi}
\psi_n(\bz)= \exp(n \, \pi \, a \, \h) \, \bz \exp(- n \, \pi \, a \, \h)=
\Sigma ^{\sigma}_{\exp(n \pi a \h)} (\bz) \qquad .
\end{equation}
The orbit of a point $\bz$ of $AdS_3$
obtained by the  bi-action transformations will be denoted by
${\cal O}_{\bz}$:
\begin{equation}
\label{OZ}
{\cal O}_{\bz} = \Sigma^{\sigma}_G (\bz)
:= \{g\,\bz\,\sigma(g^{-1})|g\in \SL\} \qquad .
\end{equation}
Using the matrix parametrization (\ref{matbz}) of the points of $AdS_3$,
we see that the surface ${\cal O}_{\be}$ corresponds to the set of $\SL$
matrices $\bz$ with $x(\bz)=0$. More generally, we have:
\begin{equation}
{\cal O}_{\bz_0}=\{ \bz \, | \, x(\bz)=x(\bz_0) \} \qquad .
\end{equation}
These orbits are
stable with respect to the action of the
one parameter BHTZ subgroup (\ref{defpsi}).
Let us determine the subgroup $Stab(\bz)$ (connected to
the identity) of
$\SL$ leaving the point
$\bz$ fixed under the bi-action.
Infinitesimally, it is generated by the
$\ls$ generators $\n$ satisfying the equation:
\begin{equation}
\n\,\bz-\bz\,\sigma(\n)=0 \qquad ,
\end{equation}
whose solution is, up to a multiplicative factor:
\begin{equation}
\n = \bz\,\h-\frac 12\tr(\bz\,\h){\Id}\qquad .
\end{equation}
A vector tangent to ${\cal O}_{\bz}$, at $\bz$, can be expressed as:
\begin{equation}
\label{vecl}
  \Lambda _{\bz}={\LL}\,\bz- \bz \, \sigma(\LL) \qquad \mbox{\rm
with}\qquad {\LL} \in \ls \qquad .
\end{equation}
and an orthogonal vector may be written as:
\begin{equation}
\label{Nz}
  N_{\bz}={\overline \n} (\bz) = \n \bz \qquad ;
\end{equation}
this can be verified
using the cyclic property of the trace.
This vector is everywhere spacelike on $AdS_3$,
its norm being strictly positive:
\begin{equation}
\label{normN}
\|{N}_{\bz} \|^2=1+ x(\bz)^2 \qquad ,
\end{equation}
where $x(\bz)$ is given by the matrix representation of the point $\bz$
(see eq. (\ref{matbz})).

The domain $\cal U$ that, after identification, will provide black hole
solutions, can be chosen as one connected component of the open region
of $G$ defined by:
\begin{equation}
\label{Ureg}
\|  \Xi _{\bz} \| ^2 = 2 a^2 \, \left ( t(\bz)^2-y(\bz)^2\,
\right ) >0 \qquad ,
\end{equation}
with $y(\bz)$ and $t(\bz)$ defined as $x(\bz)$. Note that, as the identity
element $\be$ is fixed by the BHTZ subgroup, it belongs to
the boundary of the open domain $\cal U$.

Normalizing the vector field ${N}_{\bz}$, we obtain a (unit) vector field
$ \nu_{\bz}$ that defines a global one-parameter group
of diffeomorphisms on $\SL$ (the flow of the vector field $\nu$):
\begin{equation}
\label{diff}
({\rho},\bz)\mapsto \phi_{\rho}(\bz)\qquad \mbox{\rm with}\qquad \frac
{d\phi_{\rho}(\bz)}{d \rho}={\nu}_{\phi_{\rho}(\bz)}\quad
\mbox{\rm and }\quad \phi_0(\bz)=\bz \qquad .
\end{equation}
This diffeomorphism commutes with the bi-action. To prove this, it is
  sufficient to show that
the flow $\phi$ of the vector field $\nu$ and the action
$\Sigma ^{\sigma}$ (eq. (\ref{tau})) satisfy:
\begin{equation}
\label{condcom}
\Sigma^{\sigma}_g \, \phi _\rho \, \Sigma^{\sigma}_{g^{-1}} = \phi 
_\rho \qquad \forall
g \in G \qquad ,
\end{equation}
which is equivalent to ${\Sigma ^{\sigma} _g}_* \nu = \nu$, where $*$ denotes
the differential mapping. Observing that at a given $\bz \in
AdS_3$ the vector  $ \nu_{\bz}$ can be viewed as
the right translated of a multiple of the generator $\n$ belonging
to the stabilizer algebra of $\bz$ with respect to the bi-action
(see eq. (\ref{Nz})), the condition (\ref{condcom})
reduces to:
\begin{equation}
g \, \left ( \, Stab_G(\bz)\, \right ) g^{-1}
\subseteq Stab_G(\Sigma^{\sigma} _g (\bz)) \qquad ,
\end{equation}
which holds trivially. The group $G$ being connected, this
  can also be checked
by verifying that the Lie bracket of the vector
fields (eqs (\ref{diff}), (\ref{vecl})):
\begin{equation}
[\nu,\,\Lambda]_{\bz} =
\left .\frac d{ds}\frac
d{d\rho}\phi_{\rho}(\Sigma^{\sigma}_{\exp[s\,{\LL}]}[\bz])\right|_{\rho=0,\ 
s=0} -
\left .\frac d{d\rho}\frac
d{ds}\Sigma^{\sigma}_{\exp[s\,{\LL}]}[\phi_{\rho}(\bz)]\right|_{\rho=0,\
s=0} = 0\qquad .
\end{equation}
vanishes.

Now we dispose of all geometric ingredients necessary to build a global metric
on BTZ black holes. We have a foliation -~a trivial fibration~- of
$AdS_3$ whose sheets are the $\SL$ invariant  ${\cal O}_{\bz}$ orbits
stable with respect to the BHTZ subgroup action given in eq. (\ref{zet}).
All these orbits are diffeomorphic, with the diffeomorphism being realized
by the normal flow defined in eq. (\ref{diff}). This essentially reduces
the geometrical description of $AdS_3$ from three to two dimensions.
Accordingly, we will obtain an expression of the $AdS_3$ metric of the type:
\begin{equation}
d \rho^2 + f^2(\rho) \, d s_{\rho=0} ^2 \qquad ,
\end{equation}
where $f^2(\rho) \, d s_{\rho=0} ^2$ is the pullback
metric on the sheet ${\cal O}_{\be}$ by the diffeomorphism
$\phi_\rho$. All $\SL$ invariant Lorentzian metrics
are proportional to the metric of a unit time-like hyperboloid in a
3-dimensional Minkowskian space.
Explicitly, the construction of the global metric goes as follows.

Consider the orbit ${\cal O}_{\be}$ containing
the identity
  element $\be$, defined in eq. (\ref{OZ}).
From the solution:
\begin{equation}
\label{phirho}
\phi_{\rho}(\be)=  \exp (\rho \, \h)
\end{equation}
of the differential equation (\ref{diff}),
with point $\be$ as initial condition, we obtain the
solution from any other
point $g\,\sigma(g^{-1})$ of ${\cal O}_{\be}$ as
$g\,\phi_{\rho}(\be)\, \sigma(g^{-1})$. This
can be checked from the explicit expressions of $\sigma$
(eq. (\ref{defsig})) and
$\nu $ (eqs (\ref{Nz}, \ref{normN})).
Note also that two $\SL$ elements $g$ and $g'$ lead to the same point of
the orbit
${\cal O}_{\be}$ if and only if
the product $g^{-1}g'$ commutes with
$\h$, i.e. if and only if $g' = g\, h$ with $h=\pm \exp[\alpha \h]$.
Furthermore, on $\ls$ (which is isomorphic to  3-d Minkowskian space)
the Lorentz group SO(1,2) acts on an element ${\LL}$ as
$\mbox{Ad}(G)\,{\LL}=\{g\, {\LL} g^{-1}|g\in \SL\}$.
On the other hand, the adjoint orbit of $\h$ in $\ls$ under the 
Lorentz group, the hyperboloid
${\cal H}_{\h}$, is diffeomorphic to the orbit ${\cal O}_{\be}$ via:
\begin{equation}
\iota \ : \quad {\cal H}_{\h} \rightarrow {\cal O}_{\be} \ : \
Ad(g) \h \mapsto \Sigma^{\sigma} _g (\be) \qquad .
\end{equation}
Indeed, this map defines a $G$-equivariant bijection such that
$Stab_{Ad_G}(\h)= Stab_{\Sigma^{\sigma}_G}(\be)= \pm \exp(\mathbb{R} \h)$.

The coordinate system $(\tau,\theta)$ on ${\cal O}_{\be}$ defined by:
\begin{equation}
(\tau,\theta) \rightarrow \Sigma ^{\sigma} _{\exp[\frac \theta 2 \h]
\exp[(\frac \tau 2 + \frac \pi 4)\t]} (\iota \s) \quad ,
\quad {\rm with} \quad \s=\e+\f \quad ,  
\end{equation}
is well adapted to the action of the BHTZ group $\ZZ \simeq \exp(\ZZ \h)$.
In matrix form, it reads:
\begin{eqnarray}
&&\left(\exp[\frac \theta 2 \h]\exp[\frac \tau 2 \t]\h
\,\exp[-\frac \tau 2 \t]\exp[- \frac \theta 2
\h]\right)\h \nonumber \\  \nonumber  \\
&&=\left(\begin{array}{cc}
\cos(  \tau)&\sin( \tau)\exp(\theta)\\
-\sin( \tau)\exp(-\theta)&\cos( \tau)
\end{array}\right)\qquad .
\end{eqnarray}
These coordinates cover only the intersection of
${\cal O}_{\be}$ with the interior of the null cones whose vertex
are located at the identity element $(\be\equiv \Id)$ and its
opposite ($-\Id$). Accordingly, it is well behaved on the domain defined by
the intersection of the region $\cal U$, eq.
(\ref{Ureg}), and ${\cal O}_{\be}$. In terms of these
coordinates, the metric reads locally as:
\begin{equation}
ds_{\rho=0}^2=-d\tau^2 + \sin(\tau)^2\,d\theta^2\qquad {\rm with} \qquad
0<\tau<\pi \ ,\ -\infty<\theta<\infty \quad .
\end{equation}
This metric is clearly $\SL$-invariant.

In order to obtain global coordinates on the black hole we transport
the previous
coordinate patch using the diffeomorphism $\phi_{\rho}$
defined in eqs  (\ref{diff}) and (\ref{phirho}), with $-\infty<\rho<\infty$.  This
leads to the expression for $\bz$:
\begin{equation}
\bz(\tau,\theta, \rho)=\left(\begin{array}{ll}
\sinh(\frac \rho 2)+ \cosh(\frac \rho 2)\,\cos(\tau)&
\exp(\theta)\,\cosh(\frac \rho 2)\,\sin(\tau)\\
-\exp(-\theta)\,\cosh(\frac \rho 2)\,\sin(\tau)&
-\sinh(\frac \rho 2)+ \cosh(\frac \rho 2)\,\cos(\tau)
\end{array}
\right)\qquad ,
\end{equation}
which parametrizes the whole domain $\cal U$
(see eq.(\ref{Ureg})). We therefore obtain the metric expression:
\begin{equation}
\label{globmet}
ds^2=\frac 14 d\rho^2+\cosh^2(\frac \rho 2)(-d\tau^2 +
\sin^2(\tau)\,d\theta^2)\qquad .
\end{equation}
The BHTZ subgroup action reads:
\begin{equation}
(\tau,\theta, \rho) \mapsto (\tau,\theta + 2 n\, \pi \, \sqrt{M}, \rho) \qquad ,
\qquad n \in \ZZ  \qquad ,
\end{equation}
and therefore the black hole is globally modelled by the Lorentzian manifold:
\begin{equation}
\label{manif}
\left ( \quad ]0,\pi[ \times S^1 \times \mathbb{R} \, , \, ds^2
\quad \right ) \qquad .
\end{equation}
In particular, we recover the $S^1 \times \mathbb{R}^2$ topology described
in \cite{BTZ}.

\section{Poisson structure}
In this section, we prove that a BHTZ non-rotating massive black
hole is canonically endowed with a regular Poisson structure.
The latter admits as characteristic foliation the one constituted by the
orbits of the external bi-action $\Sigma^\sigma$.

We have seen in the previous sections that a BTZ non-rotating massive black
hole is globally isometric to the manifold $(]0,\pi[\times S^{1}\times
\RR,ds^2)$ (cf. eq. (\ref{manif})). The submanifold $]0,\pi[\times 
S^{1} \times \{0\}$ is covered by an open domain of a totally
geodesic Lorentzian submanifold of $G=\mbox{AdS}_3$ namely the
$\Sigma^\sigma_G$-orbit $\co_{\be}$ through the neutral element. As a
$G$-homogeneous space this orbit is isomorphic to the coset space $G/H$ where
$H=\pm\exp(\RR \h)$. Every non zero constant multiple of the area element on
$G/H$ defines a $G$-invariant symplectic structure on this space.
Restricted to the above mentioned open domain and then projected, the
latter defines on $]0,\pi[\times S^{1}\times\{0\}$ a symplectic form
denoted by $\omega$. This can be described more intrinsically in terms of
para-complex structures \cite{KANEYUKI,BIELIAVSKY} as follows.  Consider the
Lorentzian homogeneous space $G/H$.  Denote by $\eta$ the
$G$-invariant Lorentzian structure on $G/H$ obtained by projecting the
Killing metric $\beta$ (\ref{metric}).  Denote by $o:=[\be]$ the class
of the neutral element $\be$.  At point $o$, the endomorphism
\begin{equation}
J_o:T_o(G/H)\to T_o(G/H):X\mapsto\frac{1}{2}[\h,X],
\end{equation}
commutes with the action of the isotropy group $H$. It can therefore
be consistently transported to every point of $G/H$ by the action of
$G$ ($\Sigma^{\sigma}$), defining
globally on $G/H$ a smooth $G$-invariant (1,1)-tensor field $J$.
The tensor field $J$ is called the canonical {\bf para-complex structure}
on $G/H$ and enjoys the following properties:
\begin{enumerate}
\item[(i)] $\forall x\in G/H:\,J_x^2=\mbox{id}_{T_x(G/H)}$;
\item[(ii)] $J_x$ is skewsymmetric with respect to
the Lorentzian scalar product
$\eta_x$
on $T_x(G/H)$;
\item[(iii)] the $G$-invariant 2-form $\omega:=\eta(J\,.\, ,\, .\,)$ on
$G/H$ is symplectic.
\end{enumerate}

Each of the submanifolds $]0,\pi[\times S^{1}\times\{\rho\}$ of the 
BTZ-space $(]0,\pi[\times S^{1}\times
\RR,ds^2)$ is canonically endowed with a para-complex structure $J^\rho$.
Extending $J^\rho$ by $0$ on the $\partial_\rho$-direction, one
obtains a (skewsymmetric) smooth (1,1)-tensor field $\cj$ on the
BTZ-space.  Let $\cf$ denote the foliation of the BTZ-space whose
leaves are the above submanifolds.  Then the leafwise symplectic
2-form
\begin{equation}
\omega^{\cf}:=ds^2(\cj \,.\,,\,.\,)
\end{equation}
defines a Poisson structure on the BTZ-space whose characteristic
foliation is $\cf$.

In coordinates $(\tau,\theta,\rho)\quad 0<\tau<\pi$ on the BTZ space,
the Poisson structure corresponding to $\omega^{\cal F}$ reads
\begin{equation}
\{\,,\,\}=\frac{1}{\cosh^{2}(\frac{\rho}{2})\sin(\tau)}\partial_\tau\wedge\partial_\theta.
\end{equation}

\section{Discussion}

The global BTZ metric obtained in this paper (eq. (\ref{globmet}))
provides an explicit geometric description of BTZ space as a
succession of hyperbolic planes quotiented by a one-parameter
subgroup, leading to a Riemannian surface whose moduli evolves in a
time dependent way from an initial to a final singularity (a ``cylinder'' built
from identifications performed
on a Lobachevsky plane, whose radius starts from zero,
grows to a maximum and then decreases again to zero).
Moreover let us also notice that the surface $\rho=0$ of this black 
hole universe constitutes a minimal
surface and accordingly a solution of the classical string equations 
\cite{BaPe} representing a closed string
winded around the throat of the black hole. This configuration is 
obviously stable. We shall present its
spectrum of perturbations in a forthcoming publication \cite{BRS}.

Furthermore, the group theoretical developments performed in this paper
offer several perspectives. First of all, the definition of a regular
Poisson structure on BTZ space opens the possibility of quantization.
It should be noted that
the action of the integers on $AdS_3$ leading to the BTZ-space is
generally seen as the unique isometric action extending to $AdS_3$ the
action of a Fuchsian group $\Gamma=\mathbb Z$ on a totally geodesic
Riemannian submanifold ${\bf D}\subset AdS_3$ isometric to the hyperbolic
disc \cite{Brill}. In this case, the extended action is interior, hence it is
($\mbox{Iso}(\mbox{AdS}_3)$-conjugated to) the bi-action 
($\Sigma^{\mbox{id}}$) by conjugation:
$G\times G\to G:(g,x)\mapsto gxg^{-1}$. The associated 
Poisson-structure on $AdS_3$
is non-regular in this case. Indeed, near the neutral element, this
Poisson structure appears to be (locally) isomorphic to the canonical linear
one on the dual of the Lie algebra $\cg^\star$ whose characteristic 
foliation is given by the
coadjoint orbits (for precise definitions see e.g. \cite{VAISMAN}).

Another perspective is the
   generalization of the approach
   to perform $AdS_3$ quotients with Fuchsian groups and hence to
   have a way to handle BTZ solutions on spaces with more complex
   topologies, whose spatial infinity surface has, for example, several
   disconnected components. The existence of such spaces has already been
  suggested earlier \cite{Brill}.

The search for BTZ solutions in spaces with different topologies lies
in the framework of the quest for the origin of the black hole
   entropy. This entropy seems due to a macroscopically large number of physical
   degrees of freedom, as indicated by the value of the central charge first
   computed in \cite{BH}. Up to now, these degrees of freedom have not
   been seen at the classical level. However, the opposite
has not been shown either. This paper participates to the effort of
determining whether or not these degrees of freedom could be hidden in the
existence of a
macroscopically large number of different BTZ topologies,
obtained by moding out
$AdS_3$ by different Fuchsian groups.  
\section*{Acknowledgments} P.B.
is supported by the Communaut\'e Fran\c caise de Belgique, through an
Action de Recherche Concert\'ee de la Direction de la Recherche
Scientifique. MR and PhS acknowledge support from the Fonds National de la
Recherche Scientifique (F.R.F.C. contracts).

\newpage

\section{Figure captions}

{\bf Figure 1.} Visualization of the exponential mapping $\exp:
\ls \to\SL$ restricted to future time-like and null adjoint orbits.
The target space $\SL$ is parametrized via the coordinates
$(r,\lambda,\varphi)$ (cf. eq. \ref{he}).  Note that, under $\exp$,
adjoint orbits ($B,B_{0}$) correspond to orbits (${\cal B},{\cal B}_{0}$) for
the action by conjugation ($\Sigma^{\mbox{id}}$) in $G=\SL$, that is to
the space-like surfaces considered in \cite{Brill}. Those are hyperbolic
planes when endowed with the restricted metric. The hyperbolic plane
${\cal B}_0$ is totally geodesic.
In the ``3-D  Penrose
diagram", i.e. the part of the static Einstein cylindrical space
limited by $r\leq \pi/2$,
the point $\be=\Id$ has coordinates $(0,0,0)$ and those of $-\Id$ are $(0,0,\pi)$.
In
this cylinder the surfaces of constant $\theta$ coordinate are given by
$\sin r \, \sin \varphi=\tanh \theta\,\sin \lambda$.  The exponential
map sends the half-cone $L_+$ issued from the origin of $\ls$ on the
future half cone ${\cal L}_+$ whose vertex is on the point
corresponding to the identity element in $\SL$.  The sheet of the
hyperboloid $L_+$ is mapped on the past null cone issued from the
$AdS_3$ point corresponding to the element $-\Id$ of $\SL$.  Clearly
${\cal L}_-$ and ${\cal L}_+$ constitute the past and future Cauchy
horizons of ${\cal B}_0$.\bigskip

{\bf Figure 2.} Visualization in coordinates
$(r,\lambda,\varphi)$ of the trivial fibration of $AdS_3$ by the orbits
of the twisted action $\Sigma^\sigma$.  The orbit ${\cal
O}_{\exp(\rho\h)}$ is given by $x(\bz)=\sinh (\rho/2)$ i.e. $\tan
r\,\cos \varphi=\sinh (\rho/2)$.  The mapping $\iota$ sends the
hyperboloid ${\cal H}_{\h}$ on the orbit ${\cal O}_{\bf e}$ of the
identity element of $\SL$.  More generally, $\ls$ time-like
hyperboloids are in correspondence with $\Sigma^{\sigma}$-orbits
($\Sigma^\sigma_g(\exp(\rho H))=\iota\mbox{Ad}(g)(\rho
H)\quad\rho>0$).  When these hyperboloids shrink to the null cone
${\cal C}_{-\infty}$, the corresponding orbits go to an half of the
conformal infinity of $AdS_3$, the surface $r=\pi/2
,\pi/2\leq\varphi\leq 3\,\pi/2$ in the Einstein space. On the other
hand, when the hyperboloids blow up to infinity, the second half of the
conformal infinity of $AdS_3$ is reached in the limit. \bigskip

{\bf Figure 3.} Representation, on the same part of
the Einstein cylinder
as in Figs {\bf 1} and
{\bf 2}, of the
intersection of sections of constant time $\lambda$ (in red) with
surfaces of constant $\theta$ (in yellow:
$\theta =0$, in green: $\theta=0.55$, in blue: $\theta= 2.2$ and in
black: $\theta = -4.4$). The portion of the
space-time included between two such surfaces $\theta_0$ and
$\theta_1$ corresponds to a spinless BTZ black hole
of a given mass $M=(\theta_1-\theta_0)^2/\pi^2$ in geometrical units.
The initial and final singularities
correspond to the intersections (the red straight lines in the
surfaces $\lambda=0$ and $\lambda=\pi$)  of all
the surfaces of constant
$\theta$.

\end{document}